\documentclass[
 reprint,  %preprint,
 amsmath,amssymb,aps,showpacs,superscriptaddress,prl%,dvipdfm   % nofootinbib, showkeys
]{revtex4-1}
\usepackage{graphicx}
\usepackage{booktabs}
\usepackage{dcolumn}% Align table columns on decimal point
\usepackage{bm}% bold math
\usepackage{mathrsfs}
\usepackage{mathtools}
\usepackage{epstopdf}
\usepackage{units}
\usepackage{dsfont}
\usepackage{upgreek}

\begin{document}
\preprint{APS/123-QED}

\title{Steering most probable escape paths by varying relative noise intensities}

\author{Paul H. Dannenberg}
\affiliation{Duke University, Department of Physics, Box 90305
Durham, NC 27708-0305}

\author{John C. Neu}
\affiliation{Duke University, Department of Biomedical Engineering, Box 90281
Durham, NC 27708-0281}

\author{Stephen W. Teitsworth}
\affiliation{Duke University, Department of Physics, Box 90305
Durham, NC 27708-0305}
 \email{teitso@phy.duke.edu}

\newcommand*\separatrix{\includegraphics{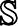}}

\date{\today}

\begin{abstract}
We demonstrate the possibility to systematically steer the most probable escape paths (MPEPs) by adjusting relative noise intensities in dynamical systems that exhibit noise-induced escape from a metastable point via a saddle point. Using a geometric minimum action approach, an asymptotic theory is developed which is broadly applicable to fast-slow systems and shows the important role played by the nullcline associated with the fast variable in locating the MPEPs.  A two-dimensional quadratic system is presented which permits analytical determination of both the MPEPs and associated action values.  Analytical predictions agree with computed MPEPs, and both are numerically confirmed by constructing prehistory distributions directly from the underlying stochastic differential equation.
\end{abstract}

\maketitle
%\tableofcontents

%======================================================================
%============================ MAIN TEXT ===============================
%======================================================================

%-------------------------- Introduction ------------------------------
The phenomenon of noise-induced escape from a metastable state occurs in a wide range of dynamical systems including chemical reactions \cite{Dykman_JCP_1994}, micromechanical oscillators \cite{Chan_PRL_2007, Chan_PRL_2008}, genetic regulatory networks \cite{Acar_NatGen_2008}, nonlinear electronic transport in semiconductor superlattices \cite{Bomze_PRL_2012},  population dynamics \cite{Khasin_PRL_2009}, epidemiological models \cite{Forgoston_Chaos_2009}, and models of cancer cell proliferation \cite{Lee_PRE_2010}.  The noise amplitude for many such systems is small, so that the system remains in the vicinity of an initial metastable point $\mathbf{x_s}$ for a long time.  However, rarely occuring configurations of the noise can drive the system out of the basin of attraction of $\mathbf{x_s}$.  Two central questions concerning noisy escape are: 1) by which paths in phase space is the system most likely to escape, and 2) what is the mean time for escape to occur?  As the noise amplitude diminishes, the paths through phase space by which escape proceeds become increasingly predictable, so that nearly all realizations of the escape path lie in a narrow tube connecting the metastable point $\mathbf{x_s}$ to a point on the boundary of its basin of attraction \cite{Chan_PRL_2008}. In such cases, one defines the most probable escape path (MPEP) as the curve that is approached by this tube as the noise amplitude tends to zero.   For near-equilibrium systems such as chemical reactions, the MPEPs generally coincide with the time-reversed saddle-node trajectories of the related deterministic system, a manifestation of detailed balance \cite{Dykman_JCP_1994,Luchinsky_RPP_1998}.

Noise-induced escape dynamics in \emph{far-from-equilibrium} systems do not generally satisfy detailed balance, so the MPEPs may differ substantially from the deterministic saddle-node trajectories \cite{Luchinsky_PRL_1999, Luchinsky_RPP_1998}.  In this paper, we obtain new and general predictions concerning noise-induced escape processes in far-from-equilibrium systems that take the form of two-dimensional fast-slow systems.  Systems of this type occur throughout the natural sciences, including, for example, models of neuron function \cite{Newby_PRL_2013,Izhikevich_2007}, population dynamics \cite{Khasin_PRL_2009}, and models of climate change \cite{Berglund_2006, Monahan_2011}; thus, the results presented here are expected to be broadly applicable.    
In far-from-equilibrium systems, it is well-known that the MPEP can be found in principle by minimizing a stochastic action functional; additionally, the \emph{mean escape time} is exponential in the corresponding minimum action value normalized by the noise intensity \cite{WentzellFreidlin_2012}. 
Over the past several years, MPEPs and their associated action values have been numerically computed for a variety of far-from-equilibrium systems.  Additionally, the generic structure of MPEPs has been analytically studied in the neighborhood of the fixed points \cite{Maier_PRL_1992}.  However, due to the complexity of the associated Euler-Lagrange equations for even the simplest systems, \emph{global} analytic expressions for the MPEPs -- which are solutions to these equations -- are generally not available.  Furthermore, there has been little work that systematically examines how escape dynamics are affected by varying relative noise amplitudes associated with the different dynamical variables. This is an important problem to consider since it is often possible to independently control one or more sources of intrinsic noise or to inject sources of external noise \cite{Chan_PRL_2007, Chan_PRL_2008}.

%MAY NEED TO PUT SOMETHING ABOUT SADDLE POINT AVOIDANCE SOMEWHERE IN THE INTRO?

%Due to the complexity of the Euler-Lagrange equations for even the simplest bistable systems, %global analytic expressions for the MPEPs are generally not available.

In this Letter, we use a recently developed geometric minimum action approach \cite{Heymann_PRL_2008} to analytically and numerically determine MPEPs and associated actions for noise-induced escape in two-dimensional fast-slow systems.  Use of the geometric approach allows us to carry out an asymptotic analysis of the obtained Euler-Lagrange equation that yields new insights into the global structure of the MPEPs.  For a specific quadratic model, we are able to analytically determine both the complete MPEP as well as the corresponding minimum action value, which provides an estimate of the mean escape time.

We begin by considering a two-dimensional fast-slow dynamical system for which the state vector $\mathbf{x}(t) = (x(t), y(t))$ evolves according to the stochastic differential equation
\begin{equation}
\label{vectorizedFastSlow}
\dot{\mathbf{x}} =  \begin{pmatrix}u(x,y)\\\mu \, v(x,y)\end{pmatrix} + \sqrt{\epsilon} \left(\begin{matrix}
\sigma_{1}&0\\ 0&\sigma_{2}
\end{matrix} \right) \begin{pmatrix}
\xi_1(t) \\ \xi_2(t)\end{pmatrix}.
\end{equation}
Here $u$ and $v$ give the deterministic flow in the $x$ and $y$ directions, respectively, and $0 < \mu  \ll 1$ implies slow dynamics in the $y$-direction.  We include a Gaussian noise source $\boldsymbol{\xi}(t) = \left(\xi_1(t), \xi_2(t)\right)$ whose components each have zero mean and unit standard deviation, and are delta-correlated, i.e., $\langle \xi_i(t)\, \xi_j(t')\rangle = \delta_{ij}\delta(t - t')$; the overall scale of noise is set by the small parameter $\epsilon \ll 1$.  Additionally, the noise amplitude tensor $\sigma_{ij} = \sigma_i \delta_{ij}$ is diagonal and state-independent, but with diagonal components that can be varied relative to one another, as characterized by a noise amplitude ratio $r:= \sigma_2/\sigma_1$.
 %$\boldsymbol{\xi}(t) = \left(\xi_0(t), \ldots, \xi_n(t)\right)$
%Refer to $\sigma$ then say $\sigma_{ij} = sigma_i\delta_{ij}$

\begin{figure}[h!]
\includegraphics[width=1\columnwidth]{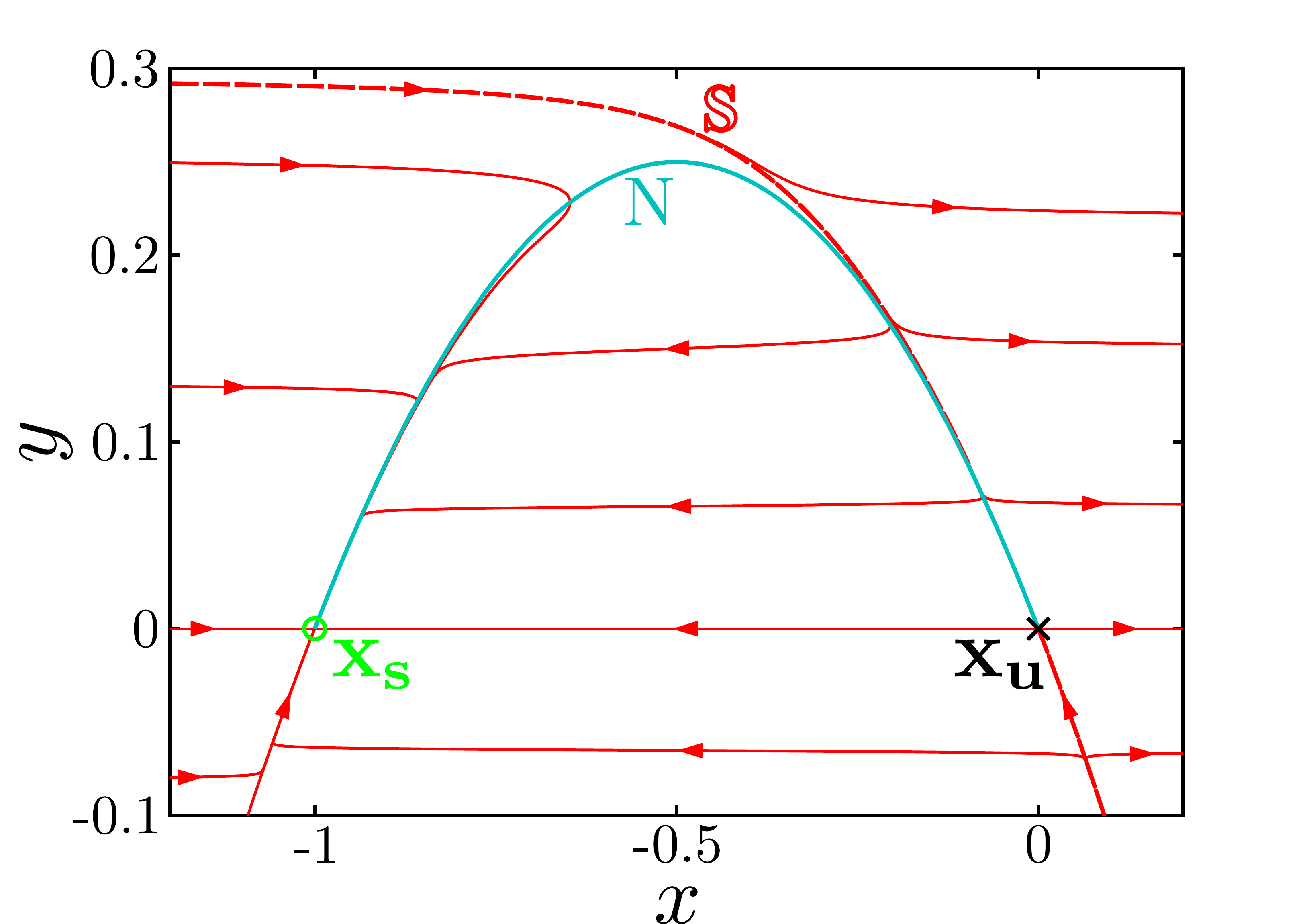}
\caption{
\label{QuadraticPP}
(color online) The phase portrait $\dot{\mathbf{x}} = (u(\mathbf{x}), \mu \, v(\mathbf{x}))^\intercal$ with $u$ and $v$ given in \eqref{dynamicalSystem}. Here, $\mathbf{x_s}$, $\mathbf{x_u}$, N and \protect\separatrix\space correspond, respectively, to the metastable point, the saddle point, the vertical nullcline (the curve along which $\dot{x} = 0$), and the separatrix.
}
\end{figure}

It is illuminating to consider a specific realization of the above fast-slow system for analysis and numerical simulation. Here we use a quadratic flow,
\begin{subequations}
\label{dynamicalSystem}
\begin{align}
u(x, y) &= x + x^2 + y \label{dynamicalSystemA}\\
v(x, y) &= -y,
\end{align}
\end{subequations}
a phase portrait of which is shown in Fig.~\ref{QuadraticPP} for $\mu = 0.01$. We focus on the MPEP that emanates from the metastable  point at $\mathbf{x_s} = (-1,0)$ and terminates at the saddle point $\mathbf{x_u}=(0,0)$. The basin boundary of $\mathbf{x_s}$ is the separatrix, \separatrix.  We also note that this system is non-gradient and hence does not generally satisfy detailed balance, since $ \sigma_2 \partial u /\partial y \neq \sigma_1 \partial v /\partial x$, except when $\sigma_2 = 0$.   % Switched citation.

To determine the MPEPs, we utilize a geometric minimum action method (gMAM), recently developed by Heymann and Vanden-Eijnden \cite{Heymann_PRL_2008, Heymann_CPAM_2008}. Generally, this method can be used to treat escape from a metastable point via a saddle point in the small noise limit for $n$-dimensional systems with dynamical variables $\mathbf{x} = (x_0, \ldots, x_n)$ that evolve according to a stochastic differential equation of the form
\begin{equation}
\label{GeneralSystem}
\dot{\mathbf{x}}(t) = \mathbf{F}(\mathbf{x}(t)) + \sqrt{\epsilon} \,\sigma(\mathbf x(t)) \,\boldsymbol{\xi}(t),
\end{equation}
where $\mathbf{F(\mathbf{x})}$ is the deterministic drift field and $\sigma(\mathbf x)$ is the noise amplitude tensor which may be state-dependent.   One considers parametric curves of the form $\mathbf{X}(s) = (X_0(s), \ldots, X_n(s))$, with $s \in [0, T]$, such that $\mathbf{X}(0) = \mathbf{x_s}$ and $\mathbf{X}(T)$ lies on the basin boundary associated with $\mathbf{x_s}$; here $s$ denotes an arclength-like parametrization. For systems with a metastable state $\mathbf{x_s}$, the basin boundary of which contains a saddle point $\mathbf{x_u}$, the MPEP generally terminates at $\mathbf{x_u}$, so that $\mathbf{X}(T) = \mathbf{x_u}$. Then, the MPEP is the particular $\mathbf{X}(s)$ that, in addition to fulfilling the two endpoint requirements, minimizes the following geometric stochastic action functional \cite{Heymann_PRL_2008, Heymann_CPAM_2008}:
\begin{equation}
\label{elegantGeoAction}
S\left[\mathbf{X}(s)\right] = \frac{1}{2} \int_C \left\{\sqrt{\left(A^{-1}\mathbf{F}\right)\cdot\mathbf{F}} \mathrm{d}s - \left(A^{-1}\mathbf{F}\right)\cdot\mathrm{d}\mathbf{X} \right\},
\end{equation}
where $C$ is the curve traced by $\mathbf{X}(s)$ joining $\mathbf{x_s}$ to $\mathbf{x_u}$, $\mathrm{d}s := \left(\mathrm{d}\mathbf{X} \cdot A^{-1} \mathrm{d}\mathbf{X}\right)^{1/2}$, and we have introduced the noise intensity tensor $A\left(\mathbf{X}\right) := \sigma(\mathbf{X})^\intercal\sigma(\mathbf{X})$. 
%The problem of finding the MPEP has now been reduced to finding the particular arclength parameterized curve $\mathbf{X}(s)$ that connects the two points $\mathbf{x_s}$ and $\mathbf{x_u}$ and minimizes the geometric action $S$ given in \eqref{elegantGeoAction}.  
Furthermore, the minimizing value of this geometric action, $S_{\min}$, provides an estimation of the time for the system to escape the attractor, $\mathbf{x_s}$, due to noise; specifically, the mean escape time is proportional to $\exp\left[S_{\min}/\epsilon\right]$ \cite{WentzellFreidlin_2012}. From the structure of the integrand in \eqref{elegantGeoAction}, one generally expects that the MPEP and associated action value will change as the elements of noise intensity $A$ are adjusted relative to one another.

%$\mathrm{d}\mathbf{X} = \dot{\mathbf{X}}\,\mathrm{d}\tau$ and $\mathrm{d}s = (A^{-1}\dot{\mathbf{X}}\cdot \dot{\mathbf{X}})^{1/2} \,\,\mathrm{d}\tau$
%For systems with a metastable state $\mathbf{x_s}$, the basin boundary of which contains a saddle point $\mathbf{x_u}$, the action \eqref{TimeAction} generally achieves a minimum when the curve $\mathbf{X}(\tau)$ terminates at $\mathbf{x_u}$, so that $\mathbf{X}(T) = \mathbf{x_u}$.

Returning to the problem of determing the MPEPs for the two-dimensional system \eqref{vectorizedFastSlow}, we first examine the case $r \searrow 0$ with $\mu$ fixed, i.e., the $y$ component of noise $\sigma_2$ is insignificant. In this limit, the MPEP must follow the $x$ axis since neither the deterministic flow nor the noise can perturb it away from this path.  In the opposite limiting case, where the noise acts only in the $y$ direction, we expect the vertical nullcline -- i.e., the curve given by setting $u(x,y) = 0$ in \eqref{dynamicalSystemA}, and denoted by N in Fig.~\ref{QuadraticPP} -- to play a central role. For escape to occur, noise must drive the system against the deterministic flow. In the vicinity of the vertical nullcline, this flow's $x$ component is small. Therefore, a trajectory that escapes by closely following the vertical nullcline need only overcome the deterministic flow in the $y$ direction which, due to  $\mu \ll 1$, is likewise small. In fact, when $\mu = 0$, we can see from \eqref{elegantGeoAction} that the action, $S$, will be zero for a trajectory along the vertical nullcline. Since the action is always non-negative, this must be the action minimizing path.

\begin{figure}[h!]
\includegraphics[width=1\columnwidth]{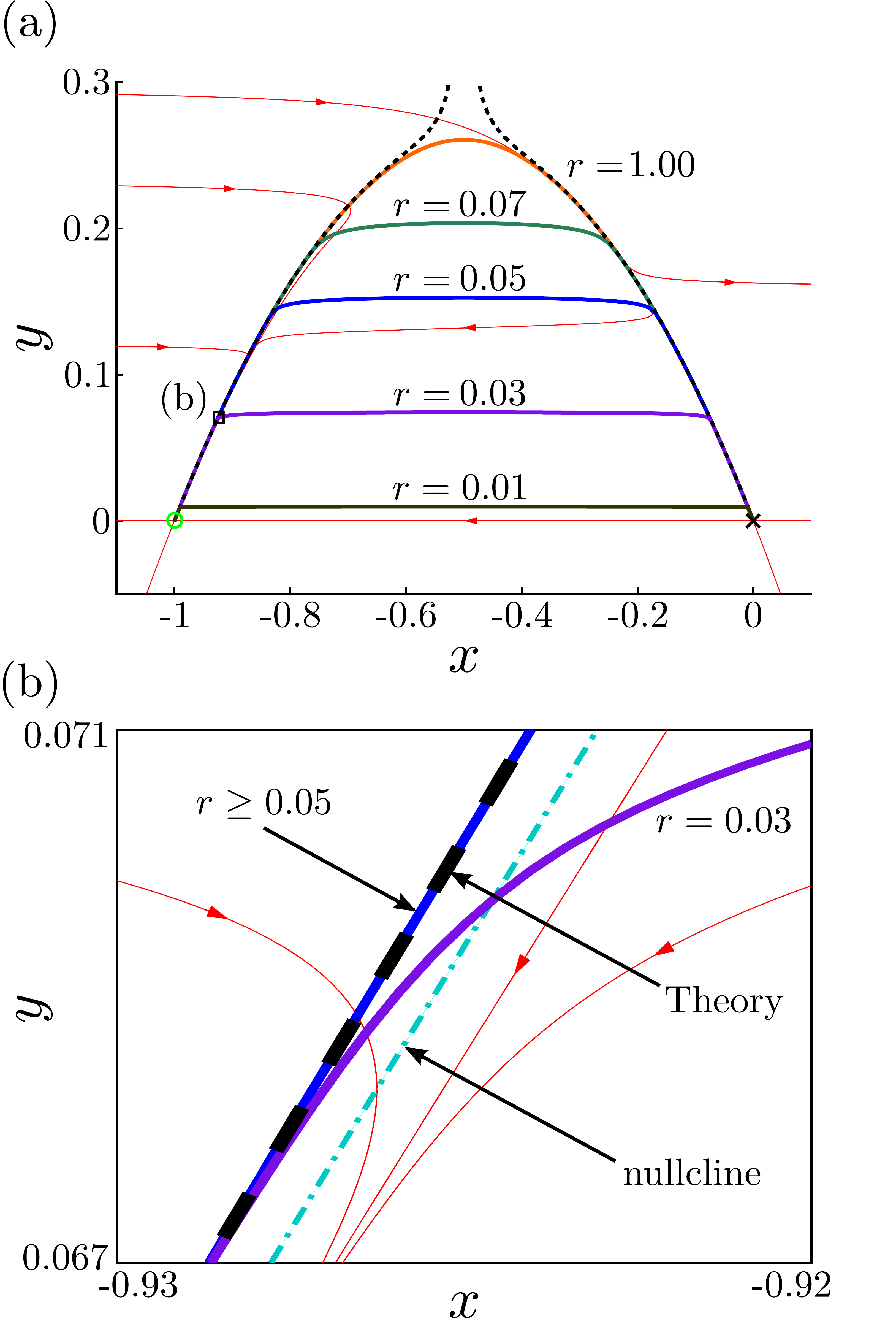}
\caption{
\label{CombinedZoomedandMPEPs}
(color online) (a) Sample gMAM computed MPEPs for various values of $r$ (thick solid curves). The analytically predicted MPEP (see \eqref{asympTheory} below) for large $r$ is shown as the dashed black line. (b) Blowup of the boxed region in figure (a) including the vertical nullcline (cyan).
}
\end{figure}

The results obtained by gMAM for several different values of relative noise strength $r$ are shown in Fig.~\ref{CombinedZoomedandMPEPs}. The two limiting cases of $r$ agree with qualitative expectations, namely, as $r$ becomes smaller the MPEP moves closer to the $x$ axis, while for large $r$ it closely follows the vertical nullcline (e.g., see the MPEP for $r=1$ which is indistinguishable from the vertical nullcline on the scale shown in Fig.~2(a)).  However, for intermediate values of $r$, rather than smoothly interpolating between the two limiting cases, the MPEP displays a segmented structure, such that it leaves the metastable point along the vertical nullcline until reaching a critical value of $y$ when it abruptly turns and follows a trajectory segment that is almost parallel to the $x$ axis, e.g., see computed MPEPs for $r = 0.01, 0.03, 0.05, 0.07$.  It should be noted that this segmented structure persists in the limit $\mu \searrow 0$, but does \emph{not} coincide with the $\mu = 0$ MPEP which follows the vertical nullcline over its entire length; similar behavior has been found in a population dynamics model \cite{Khasin_PRL_2009}. 
%We stress that the observed MPEP behaviors are not only features associated with the two-dimensional quadratic model \eqref{dynamicalSystem}, but more general.  For example, we have observed similar behavior in both two- and three-dimensional models of electrical conduction in tunnel diodes \cite{Dannenberg_Longpaper_2013}.

%We also note that for small $r$, the behavior of the MPEP is strikingly different. In this scenario the MPEP traces our analytical prediction just above the nullcline as it leaves the stable point $\mathbf{x_s}$. Similarly, as it approaches the saddle point, $\mathbf{x_u}$, it once again follows the curve defined by \eqref{asympTheory}. This section of the MPEP, for which we have found an analytical expression, is of particular importance in the phenomenon of saddle point avoidance, \cite{Luchinsky_PRL_1998}, where the shape of the MPEP results in a skewed exit location distribution along the separatrix for finite noise strengths. However, in the case of small $r$, \eqref{asympTheory} does not give good agreement over the entire MPEP. Instead we see that it contains a section that peels away from the vicinity of the nullcline and follows a path parallel to the $x$ axis. This is because the Lagrangian \eqref{ReducedLagrangian} provides an inadequate description of the numerically observed dynamics when $r$ is small.
We now explain our numerical results analytically and obtain results valid for the general class of systems described by \eqref{vectorizedFastSlow}. We start by representing the family of candidate MPEP trajectories by $Y(x)$. Next, we recast \eqref{elegantGeoAction} as an integral over $x$ yielding
\begin{equation}
\label{basicAction}
S = \int_C L(Y, Y', x) \,\mathrm{d}x\\,
\end{equation}
where $L$ is an effective Lagrangian given by
\begin{equation}
\label{EffectiveLagrangian}
L = \frac{1}{\sigma_1^2} \left\{\sqrt{u^2 + \frac{\mu^2 v^2}{r^2}}\sqrt{1 + \frac{Y'^2}{r^2}} - u - \frac{\mu v Y'}{r^2}\right\}.
\end{equation}
Solving the Euler-Lagrange equation associated with this effective Lagrangian gives the particular curve that minimizes the action, $S$. From \eqref{EffectiveLagrangian}, we see explicitly that varying $r$ will alter the trajectory of the MPEP. By contrast, scaling both $\sigma_1$ and $\sigma_2$ by identical multiplicative factors has no effect.

Firstly, we consider the case of relatively large $y$ noise component, i.e., $r$ large. To proceed, we asymptotically approximate \eqref{EffectiveLagrangian} in the limit of large $r$ and small $\mu$. Neglecting higher order $\mu$ terms, we have
\begin{equation}
\label{ReducedLagrangian}
L \sim \frac{u}{2 \sigma_2^2} \left(Y' - \frac{\mu v}{u} \right)^2.
\end{equation}%You may want to justify the fact that we have used |u| = u here.%
As discussed above, we expect the structure of the MPEP to lie close to the vertical nullcline. Furthermore, the two should coincide exactly when $\mu = 0$. We therefore conjecture that, in the limit of small $\mu$, the first portion of the MPEP has asymptotic structure
\begin{equation}
\label{HoverMPEP}
Y(x) \sim n(x) + \mu \, Y_1(x),
\end{equation}
where $Y_1(x)$ is a perturbation from the vertical nullcline, $n(x)$. Inserting \eqref{HoverMPEP} into \eqref{ReducedLagrangian} and approximating the flow in the vicinity of the nullcline using a Taylor expansion, we obtain
\begin{equation}
\label{LtoMinimize}
L \sim \frac{\mu}{2 \sigma_2^2} u_y(x,n)  \left(n' + \left(\frac{|v|}{u_y}\right)(x,n)\frac{1}{Y_1}\right)^2\,Y_1.
\end{equation}
Now, we can find the particular perturbation $Y_1(x)$ which minimizes $L$ by solving the Euler-Lagrange equation associated with \eqref{LtoMinimize}. Inserting the obtained solution for $Y_1(x)$ back into \eqref{HoverMPEP}, we arrive at the following general expression for the asymptotic MPEP, 
\begin{equation}
\label{asympTheory}
Y(x) = n(x) + \mu \frac{1}{|n'(x)|}\left(\frac{|v|}{u_y}\right)(x,\, n(x)).
\end{equation}
This curve provides an accurate approximation of the MPEP for systems with a sufficiently large noise amplitude ratio $r$, and is shown as the dashed curve in Figs.~\ref{CombinedZoomedandMPEPs}(a) and (b). Notice that there is close agreement between \eqref{asympTheory} and the numerical results obtained by gMAM except in the vicinity of $x \approx -0.5$, where the nullcline obtains a local maximum and $n'(x) \rightarrow 0$.  
%In particular, one sees in Fig.~\ref{CombinedZoomedandMPEPs} (b) that the MPEP does indeed move parallel to and just above the nullcline.

We can also use the general Lagrangian \eqref{EffectiveLagrangian} to explain the behavior of the MPEP in the case of relatively large $x$ noise, i.e., small $r$. By discounting higher order $\mu$ terms, it is possible to show that $Y' = 0$ solves the Euler-Lagrange equation corresponding to \eqref{EffectiveLagrangian} in the small $\mu$ limit. This explains the approximately horizontal segments of the MPEPs shown in Fig.~\ref{CombinedZoomedandMPEPs}(a). However, it remains to determine the point at which the MPEP peels away from its path just above the nullcline (cf. \eqref{asympTheory}) to instead follow its horizontal segment. To determine this point, we define $(x_{\mathunderscore}, y_{*})$ and $(x_+, y_{*})$ with $x_{\mathunderscore} < x_+$ as the two points where the horizontal segment of the MPEP intersects the vertical nullcline, cf. Fig.~\ref{CombinedZoomedandMPEPs}(a). Consider now an MPEP that leaves $\mathbf{x_s} = (x_s, y_s)$. First, it traces the path close to the nullcline given by \eqref{asympTheory}. Then, at the point $(x_{\mathunderscore}, y_{*})$, it peels away from the nullcline, and approximately follows a horizontal path before reaching $(x_+, y_{*})$. Finally, it falls into the saddle point, closely following the separatrix, which gives a negligible contribution to the total action, since the separatrix is a flowline of the deterministic system. Thus, to compute the total action, only contributions from the first two sections need be included. These are the integrals of the Lagrangian along their respective paths, the first of which is described by \eqref{asympTheory} and the second by the line segment $Y(x) = y_{*}$ with $x \in [x_{\mathunderscore},x_+]$. The total action for this piecewise differentiable curve is given by
\begin{figure}[t!]
\includegraphics[width=1\columnwidth]{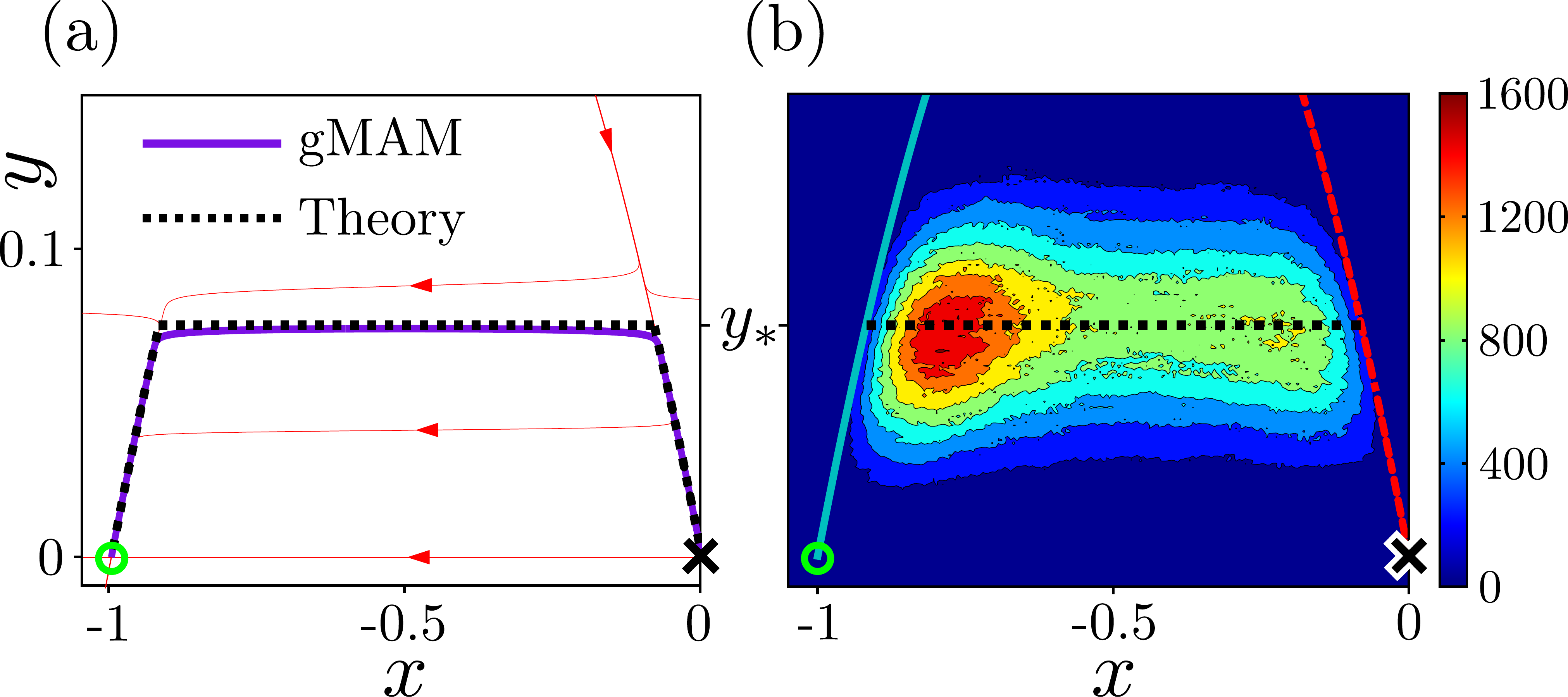}
\caption{
\label{Density}
(color online) (a) Comparison of the analytical MPEP (dashed curve) and the numerically computed MPEP (solid purple curve) for $r =0.03$. (b) Comparison of the predicted MPEP with the prehistory distribution computed directly from the underlying SDE for 20000 trials, $\mu = 0.01$, $\sigma_{1} = 1$, $\sigma_{2} = 0.03$ and $\epsilon = 0.017$. The dashed curve (red) denotes the separatrix, the solid curve (cyan) shows the left leg of the nullcline and the dotted line (black) is the horizontal portion of the predicted MPEP.
}
\end{figure}
\begin{equation}
\label{Action}
S = \frac{2 \mu}{\sigma_2^2}\int_{y_s}^{y_*} |v(x,y)| \mathrm{d}y + \frac{2}{\sigma_1^2}\int_{x_{\mathunderscore}}^{x_{+}} |u(x , y_*)| \mathrm{d}x.
\end{equation}
The MPEP and corresponding minimum action are then obtained by a one-parameter minimization of \eqref{Action} with respect to $y_*$.  A straightforward calculation shows that this point is dependent on the noise ratio $r$ such that
\begin{equation}
\label{VaryingHeight}
r = \sqrt{\frac{\mu v(x_{\mathunderscore}, y_*)}{\int_{x_{\mathunderscore}}^{x_+}u_{y}(x, y_*)\mathrm{d}x}}.
\end{equation}
Now, given a particular value of $r$, we can use this equation together with \eqref{Action} to compute both the overall shape of the MPEP and the minimizing action. As an example, for the quadratic system \eqref{dynamicalSystem} and $\mu$ small, we find that $y_*$ and the minimizing action can be analytically determined and are given, respectively, by
\begin{subequations}
\begin{align}
y_* &= \left(\frac{r}{\mu}\right)^2\left[\sqrt{4 r^4 + \mu^2} - 2 r^2\right] \label{PeelingPoint}\\
S_{\min} &\sim \frac{2  (1 + 4 y_*)^\frac{3}{2}}{3 \sigma_1^2} + \frac{2 y_*^2}{\sigma_2^2}\mu.
\end{align}
\end{subequations}
It is interesting to note that \eqref{PeelingPoint} has the correct limit as $r \searrow 0$.  Also, as can be seen from Fig.~\ref{Density}(a), the value obtained for the peeling point is consistent with the gMAM computation for $r \lesssim \mathcal{O}(\sqrt{\mu})$. However, in the limit of large $r$, there is an $\mathcal{O}(\mu^{2/3})$ discrepancy between \eqref{PeelingPoint} and the computed MPEP near the top of the trajectory at $x = -0.5$.

Analytical predictions and gMAM-computed MPEPs are confirmed by comparing them with direct simulations of the stochastic differential equation \eqref{dynamicalSystem} using a standard Euler-Maruyama method \cite{Gardiner_2009}. The \emph{prehistory distribution} \cite{Dykman_PRL_1992} for escaping trajectories at a particular noise amplitude ratio $r=0.03$ is shown in Fig.~\ref{Density}(b). This data was collected by simulating the system at low noise strength and waiting for its state to first cross the separatrix. We then traced the escape trajectory backwards in time to the instant when it \emph{last} crossed the nullcline in the neighborhood of $(x_{\mathunderscore}, y_{*})$, adding the intervening positions of the system to the density plot.  The distribution of escaping trajectories is clearly centered on the horizontal segment of the MPEP at the predicted height $y_*$, with a slight downward dip evident as it approaches the separatrix. 

In conclusion, we have demonstrated the possibility to systematically steer the most probable escape paths by adjusting relative noise intensities in dynamical systems that feature noise-induced escape from a metastable point via a saddle point. Starting from a geometric formulation of the stochastic action, we have developed an asymptotic theory, applicable to two dimensional fast-slow systems for a range of relative noise intensities, that shows the important role played by the nullcline associated with the fast variable.  In particular, the MPEP is observed to exhibit a segmented structure in which it follows close to this nullcline for at least part of its trajectory. 
%A quadratic model has also been presented which permits the determination of analytical expressions for both the MPEPs and their associated action values.  Finally, the validity of both the analytically and numerically determined MPEPs has been confirmed by direct simulation of the SDE for the quadratic system.

We thank Joshua Socolar for helpful comments.  This work was supported in part by NSF grant DMR-0804232.

%We conjecture that the tendency of the MPEP to follow close to the slow nullcline is also likely to be found in fast-slow dynamical systems with dimension greater than two \cite{Dannenberg_Longpaper_2013}
%Took this out of conclusion. I think it could appear earlier though.

\bibliographystyle{apsrev4-1.bst}
\bibliography{shortPaper3}

\end{document}